\documentstyle[11pt,aaspp4]{article}

\def\mathnew{\mathsurround=0pt}
\def\simov#1#2{\lower .5pt\vbox{\baselineskip0pt \lineskip-.5pt
        \ialign{$\mathnew#1\hfil##\hfil$\crcr#2\crcr\sim\crcr}}}
\def\simg{\mathrel{\mathpalette\simov >}}
\def\siml{\mathrel{\mathpalette\simov <}}

\received{ 5/31/1996}
\slugcomment{Submitted to Ap.J.(Lett.), 5/31/96}

\begin{document}

\title{OPTICAL AND LONG WAVELENGTH AFTERGLOW FROM \\
GAMMA-RAY BURSTS }

\author{P. M\'esz\'aros\altaffilmark{1} }
\affil{525 Davey Laboratory, Pennsylvania State University, University Park, PA 16802}
\altaffiltext{1}{Center for Gravitational Physics and Geometry, Pennsylvania
  State University}
%\authoremail{ }

\and

\author{ M.J. Rees}
\affil{Institute of Astronomy, Cambridge University, Cambridge CB3 0HA, U.K.}

\begin{abstract}

We discuss the evolution of cosmological gamma-ray burst remnants,
consisting of the cooling and expanding fireball ejecta together with 
any swept-up external matter, after the gamma-ray event. We show that 
significant optical emission is predicted which should be measurable for 
timescales of hours after the event, and in some cases radio emission
may be expected days to weeks after the event. The flux at optical, X-ray
and other long wavelengths decays as a power of time, and the initial value 
of the flux or magnitude, as well as the value of the time-decay exponent,
should help to distinguish between possible types of dissipative fireball 
models.

\end{abstract}

\section{Introduction}

Gamma-Ray Bursts (GRB) must leave behind remnants, hereafter referred to as
GRBR. In general, GRB at any plausible distance should result in 
relativistically expanding fireballs (e.g. ~\cite{me95} for a review), and the 
$\gamma$-ray emission most probably arises after the fireball becomes optically 
thin, in shocks occuring either beacuse the ejecta run into an external medium,
or because internal shocks occur in a relativistic wind. In the first type 
(a) of GRB model, the initial energy input is impulsive and the relativistic 
ejecta begin to decelerate when they have swept up external mass that is a
fraction $\sim \Gamma^{-1}$ of the ejecta mass, leading to an external and a 
reverse shock that radiate a large fraction of the initial total energy 
(the GRB event, ~\cite{rm92,mr93a,ka94,sanapi96}). However, after this initial 
burst the ejecta should still continue to expand, sweeping up an increasing 
amount of matter and slowing down.  As pointed out by ~\cite{pacrho93}, the 
evolution of such an ``old" cooling fireball should resemble the Van der Laan 
model of expanding radio sources, and they estimated the possibility of 
detecting the late radio emission from such objects. In the second type (b) of 
GRB model (\cite{rm94,paxu94}), the initial energy input continues over a 
period of time $t_w$ and the resulting relativistic fireball wind produces a 
GRB due to internal shocks in the wind itself. These objects too should leave 
behind a cooling remnant, although the dynamics of the late evolution are 
expected to be different from those of type (a) because the burst occurs at 
significantly smaller radii and the effects of the external matter are 
negligible for some time afterwards so the dynamics are different. 

In this paper we investigate the dynamical evolution of the GRBR following 
the GRB event, both for the impulsive external shock and the wind internal 
shock models. We assume that the GRB are at cosmological distances (similar 
models with somewhat different radiation characteristics can be calculated 
for galactic halo models too). Both types of cosmological GRBR produce,
in addition to decaying $\gamma$-rays, significant amounts of softer 
radiation, mostly X-rays and optical, but in some cases also radio.
Depending on the distance to the source, the duration of the main GRB event 
and the physics of the burst model, this radiation can be in many cases 
detectable with appropriate instrumentation. Based on practical considerations 
of sensitivity, telescope availability and fast response time, the best chances 
for detection at energies other than gamma-rays may be at optical wavelengths.

\section{Dynamical Evolution of GRBR}

A source at distance $D$ expanding isotropically with a relativistic bulk 
Lorentz factor $\Gamma$, which in its own frame has a comoving specific 
intensity $I'_{\nu_m}$ at comoving frequencu $\nu'_m$, will produce as 
a function of observer time $t$ an observer-frame flux at frequency $\nu_m$
\begin{equation}
F_{\nu_m} \sim \frac{c^2 t^2 \Gamma^2}{D^2} \Gamma^3 I'_{\nu_m},
\label{eq:flux}
\end{equation}
where $\nu_m=\Gamma\nu'_m$ and where the observed transverse apparent 
size $ct\Gamma$ is used (\cite{rees66}). 
In the initial phase of expansion of the fireball the internal energy density 
drops adiabatically and most of the energy is locked up in the expansion 
kinetic energy, so that $I'_{\nu_m}$ is negligible until after shocks 
dissipate some or all of the bulk kinetic energy in the GRB event. At this
point, the GRB spectrum is generally considered to be due to synchrotron or 
inverse Compton (IC) radiation from power-law electrons or pairs accelerated in 
the shocks. A break in the spectrum appears at photon energies associated 
either with synchrotron photons whose index changes below and above the 
minimum synchrotron frequency corresponding to electrons with $\gamma_{min}$, 
or else with IC photons corresponding to the same break but shifted up in 
energy by $\gamma_{min}^2$ (\cite{mrp94}, henceforth MRP94). If electrons have 
an energy distribution $\propto \gamma^{-3}$ above $\gamma_{min}$, the 
$\nu F_\nu$ photon spectrum below the break is $\propto \nu^{1}$ (or 
$\nu^{4/3}$, but we shall take $\propto \nu^1$ as a generic observed spectrum), 
and is flat above that. The GRB spectral breaks which are detected by BATSE 
generally in the 0.1-2 MeV region, e.g. ~\cite{ba93} (although they could have 
a much broader distribution,~\cite{naha96}), could be due to the synchrotron 
break (e.g. the PC model in RMP94, or~\cite{ka94,sanapi96}). Alternatively, a 
synchrotron spectral break might appear at significantly lower frequencies, 
and the similarly shaped IC scattered spectrum could provide the MeV break 
(MRP94, models FC and TC). A key quantity therefore, both for the GRB and the 
ensuing GRBR, is the comoving synchrotron intensity at the comoving synchroton 
peak frequency $\nu'_m \sim 10^6 B' \gamma_{min}^2 {\rm Hz}$, where $B'$ (G) 
is the comoving magnetic field strength,
\begin{equation}
I'_{\nu_m} \propto \frac{ n'_e B'^2 \gamma_{min}^2 \Delta R'}{B'\gamma_{min}^2 }
\propto n'_e  B' \Delta R' ~.
\label{eq:int}
\end{equation}
The GRB from impulsive fireball external shocks are expected to produce 
simultaneous X-ray and optical flashes (\cite{mr93b}, MRP94) whose duration 
is similar to that of the gamma-ray burst. Continuous wind fireball shocks 
also produce flashes of softer radiation which are contemporaneous with the 
GRB flash (\cite{pap96}). The relative amount of contemporaneous soft emission 
depends on the type of shocks, the particle acceleration process and the 
strength of the magnetic field in the acceleration region. For the GRBR, the
shock physics is also a major input in determining the comoving intensity 
$I'_{\nu_m}$ (equation ~\ref{eq:flux}), and thus the relative amounts of soft 
emission expected {\it after} the main gamma-ray and contemporaneous 
X-ray/optical flash is over. To a very large extent, however, the temporal 
behavior of the GRBR will be dominated by the evolution of $\Gamma$ and some 
of the related dynamic quantities, which are different depending on the 
initial energy input regime giving rise to the fireball.

\noindent
a) In the impulsive model, the fireball initially expands at its saturation 
bulk Lorentz factor $\Gamma\sim\eta= (E_o/M_o c^2)\sim$ constant, and upon 
interaction with an external medium it develops a blast wave moving into the 
external medium and a reverse wave moving into the ejecta, which radiate away 
a substantial fraction of the initial energy $E_o$ to give the GRB. This occurs 
at a relatively large radius $r_{dec} \sim 10^{16} (E_{51}/n_{ext})^{1/3} 
\eta_3^{-2/3} {\rm cm}$ over an observer-frame timescale $t_\gamma \sim 
t_{dec}\sim r_{dec} c^{-1} \Gamma^{-2} \sim 1~(E_{51}/n_{ext})^{1/3} 
\eta_3^{-8/3} {\rm s}$, which is in the range $1-10^3 {\rm s}$ for $\eta_3 
\sim 1-10^{-1}$ (this is the observer-frame burst duration, for an impulsive 
type of initial energy input occuring over a timescale shorter than 
$t_{dec}$). During the GRB the radiative timescales are short enough to ensure 
high radiative efficiency, but the initial rapid cooling is generally 
sufficient to ensure that the subsequent evolution of the remnant will be 
adiabatic. For such impulsive fireballs, if the external medium is 
approximately homogeneous, the GRBR will continue to evolve afterwards with 
\begin{equation}
\Gamma \propto r^{-3/2} \propto t^{-3/8}~~;~~ 
r \sim c t \Gamma^2 \propto t^{1/4}~.
\label{eq:impgamma}
\end{equation} 
Here $\Gamma$ is the Lorentz factor of the contact discontinuity 
(CD) between ejecta and external medium, and essentially also that of the 
external shock, $r$ is the distance advanced by the CD along the line of
sight (longitudinal size), and $t$ is observer-frame time.
The comoving expansion (adiabatic cooling) time is $t'_{ex} \sim 
r c^{-1}\Gamma^{-1} \propto t^{5/8}$, and the comoving radial extent of the
shell of fireball and swept-up matter is $\Delta R' \sim c t'_{ex} \propto
t^{5/8}$. 

\noindent
b) In the continuous input (wind) model, before the GRB the wind is expanding 
at its saturation average bulk Lorentz factor $\Gamma \sim \eta = 
(L/{\dot M_o} c^{2})\sim$ constant, and the primary energy input remains 
continuous (rather than being impulsive) over a wind duration timescale $t_w$ 
which characterizes the burst duration in the observer frame. Variations on 
a timescale $t_{var} < t_w$ of the bulk Lorentz factor $\Delta \Gamma \sim 
\Gamma$, e.g. due to corresponding $L$ or $\dot M$ variations, lead to internal 
shocks at relatively smaller radii $r_{dis} \sim c t_{var} \Gamma^2 \sim 
3\times 10^{11} t_{var,-3} \eta_2^2 ~{\rm cm}$, which randomize the relative 
kinetic energy of different shells of the wind and radiate away a fraction of 
order unity of the total wind kinetic energy (the GRB event; e.g. ~
\cite{rm94,paxu94,waxpi94}). During the shocks
(for $t \siml t_w$) $\Gamma$ does not change very much and may be taken as 
approximately constant. After the shocks cease ($t > t_w$), the wind remains
at $\Gamma \sim \alpha\eta \sim$ constant (with $\alpha \siml 1 \sim$ constant)
as long as deceleration by an external medium does not come into play. 
(If and when this occurs, an external shock should develop, leading to a 
slow motion version of an impulsive GRB; the remnant thereafter behaves
similarly to to case (a) above). In the range
$r_{dis} \siml r \siml r_{dec}$ (where $r_{dec} \sim 2\times 10^{17} 
(E_{51}/n_{ext})^{1/3} \theta^{-2/3} \eta_2^{-2/3}$ cm is reached after a time 
$t_{dec} \sim 10^3 (E_{51}/n_{ext})^{1/3} \theta^{-2/3} \eta_2^{-8/3} {\rm s}$) 
the wind and the remnant evolve with
\begin{equation}
\Gamma \sim \alpha \eta \sim {\rm constant}~~;~~
r \sim c t \Gamma^2 \propto t ~,
\label{eq:windgamma}
\end{equation}
where as before $r$ is the distance advanced by the CD (the front of the wind)
in observer time $t$. During the wind regime and for $t \leq t_w$, while
internal shocks continue producing the GRB light curve, the comoving density 
evolves as $n'\propto r^{-2}\propto t^{-2}$ and any comoving wind (transverse)
field as $B'\propto r^{-1}\propto t^{-1}$. However, for $r \simg r_{imp} \sim 
c t_w \Gamma^2 \sim 3\times 10^{14} t_w \eta_2^2 ~{\rm cm}$, or for times $t_w 
\siml t \siml t_{dec}$, the wind has stopped blowing while the flow can still 
have (temporarily at least) $\Gamma \sim$ constant and $r \propto t$. The
flow is now in an impulsive regime (observer-frame durations are $\simg t_w$), 
and the gas expands isotropically in its own rest frame. The 
very rapid cooling during the GRB again ensures that the subsequent evolution 
of the smoothed-out remnant wind is adiabatic. In this GRBR regime then 
$n'_e \propto r^{-3} \propto t^{-3}$, $B' \propto r^{-2} \propto t^{-2}$, and 
particle random Lorentz factors cool adiabatically as $\gamma \propto n'^{1/3} 
\propto r^{-1}\propto t^{-1}$. 

\section{GRBR Spectra from Impulsive Fireballs}

One can think of three main sub-cases which share the impulsive external shock 
dynamics described in \S 2 (a), but which, depending on the type of shock 
physics involved lead to different GRBR spectral evolution regimes.

\noindent
(a1) For the simplest impulsive model, only the forward blast wave radiates
efficiently (e.g. the reverse shock might be an inefficient particle 
accelerator or an inefficient radiator; this might be the case if the reverse 
shock were very weak, perhaps due to a very high Alfv\'en sound speed). This 
corresponds to the PC (piston) impulsive GRB model described in RMP94, 
e.g. their Figure 1c. As a specific example, this model has a $\gamma$-ray
fluence $S_\nu \sim \nu F_\nu t_\gamma \sim 10^{-7} E_{51} \theta_{-1}^{-2} 
D_{28}^{-2}$ erg cm$^{-2}$ near the BATSE threshold, where the total energy is 
$10^{51} E_{51}$ erg channeled into an angle $10^{-1}\theta_{-1}$ radians at a 
luminosity distance $10^{28} D_{28}$ cm (roughly 3 Gpc; the same object at 300
Mpc would give one of the brighter bursts).  This GRB spectrum satisfies amply 
the X-ray paucity constraint ($L_x \ll 10^{-2} L_\gamma$), being due to 
synchrotron radiation that peaks at $E_{br}\sim 0.5 \eta_3 B' \gamma_{min}^2$ 
MeV, with a $\nu F_\nu$ slope near 1 below $E_{br}$ and near 0 above it. 
The $\nu^1$ behavior below the break extends during the GRB at least down to 
optical frequencies, but becomes self-absorbed well above the radio range.  
This spectrum would be expected from electrons with an energy power law index 
$p\sim 3$ above $\gamma_{min} \sim \kappa\Gamma \sim 10^5 \eta_3\kappa_2$. 
Here $\eta=10^3\eta_3$ is the final coasting bulk Lorentz factor, $B'\sim 
10$ G is the turbulently generated comoving magnetic field strenght and 
$\gamma_{min}$ is the minimum post-shock electron random Lorentz factor, 
assuming that electrons and protons achieve a fraction $\kappa m_e/m_p$ of 
their equipartition energy (for details see MRP94). In the GRB event, the 
fireball loses on the order of its initial kinetic (total) energy, and we can 
assume adiabatic conditions for the GRBR evolution, i.e. the comoving cooling 
time rapidly becomes longer than the comoving expansion time.
The comoving intensity (\ref{eq:int}) from the external matter that
produced the initial GRB and which subsequently evolves adiabatically is
$I'_{\nu_m} \propto t^{-5/4}$, since $B'\propto V'^{-2/3},~n'_e\propto 
V'^{-1}\propto t^{-9/8},~\gamma\propto r^{-1}\propto t^{-1/4}$. Therefore
$F_{\nu_m} \propto t^2 \Gamma^5 I'_{\nu_m} \propto t^{-9/8}$ starts to 
decrease in time immediately after the GRB for this initially shocked shell 
of external matter, its peak frequency dropping as $ \nu_m\propto \Gamma B' 
\gamma^2 \propto  t^{-13/8}$, giving $F_{\nu_m} \propto \nu_m^{9/13}$. 
This adiabatically cooling leftover component from the GRB is, however,
weaker than that due to newly shocked external matter, as the ejecta continues
to advance at a steadily decreasing velocity into the external matter. For this
newly shocked matter $B',~n'_e,~\gamma$ are all three independently
$\propto \Gamma\propto t^{-3/8}$, so $I'_{\nu_m}\propto t^{-1/8}$.  As a 
result, from (\ref{eq:flux}) the observed flux at the break evolves initially
as $F_{\nu_m}\propto t^0\sim$ constant, so the flux below the break remains
constant, $F_{\nu} \propto \nu^0 \simeq$ constant as long as $\nu < \nu_m$, 
but the latter decreases in time as $\nu_m \propto t^{-3/2}$. If we take 
approximately the gamma-ray band to be at $10^{20}$ Hz and the optical band 
at $10^{15}$ Hz, $\nu_m$ drops below the optical band at $t_{opt}\sim 
3\times 10^3 t_\gamma$, where $t_\gamma \sim t_{dec}$ is the duration of the 
gamma-ray flash defined below equation (\ref{eq:flux}).
Thus, since the flux is $\propto \nu^{-1}$ above $\nu_m$
the detected {\it optical} flux from an impulsive (a1) type GRBR is
$F_{opt}\sim 10^{-27} K~D_{28}^{-2}=$ constant for $t\siml t_{opt}\sim
3\times 10^3 t_\gamma$, and $F_{opt}\sim 10^{-27} K~D_{28}^{-2} 
(t/t_{opt})^{-3/2}$ for $t \simg t_{opt}$, where flux units are erg cm$^{-2}$
s$^{-1}$ Hz$^{-1}$, $D$ is the luminosity distance and $K=E_{51} 
\theta_{-1}^{-2}$ is the burst energy normalization . The corresponding V-band 
magnitudes are $m_v =-2.5 \log(F_{opt}/4\times 10^{-20})$ or
\begin{eqnarray}
m_v & \simeq & 19 -2.5 \log K + 5 \log D_{28} ~~~~\hbox{,~for}~
t \siml 3\times 10^3 t_\gamma \;, \nonumber \\ & \\ 
m_v & \simeq & 19 - 2.5 \log K + 5 \log D_{28} + 3.75 \log(t/t_{opt}) ~~~~
\hbox{,~for}~ t \simg  3\times 10^3 t_\gamma \;, \nonumber
\label{eq:a1mag}
\end{eqnarray}
The optical brightness is constant at the level given by the first line of 
equation (\ref{eq:a1mag}) for about an hour (since for the fiducial parameters 
here $t_\gamma \sim 1~{\rm s}$). Note that at 300 Mpc ($D_{28}=10^{-1}$) 
the first line of (\ref{eq:a1mag}) would be $m_v \simeq 14 +\cdots$.
In this model, the self-absorption frequency (initially at $\sim 10^{13}$ Hz)
evolves at first $\propto t^{-1/4}$, slower than $\nu_m$, the two becoming
equal at $t_{m,ab}\sim 4\times 10^5 t_\gamma$ where $\nu_{ab}\sim 3\times 
10^{11}$ Hz. Thereafter, the frequency $\nu_{ab}\propto t^{-11/14}$
and the peak of the spectrum separating the optically thick and thin regimes
evolves as $F_{\nu_{ab}}\propto \nu_{ab}^{10/11}$ as long as the GRBR 
still expands relativistically (it reaches $\Gamma\sim 1$ after $t_{nr} 
\sim 10^8 t_\gamma$, about three years for $t_\gamma\sim 1~{\rm s}$),
and $F_{\nu_{ab}} \propto \nu_{ab}^{7/5}$ afterwards. As a result, the
radio flux in the 100 MHz band is negligible, $F_R \siml 10^{-2} K D_{28}^{-2} 
{\rm \mu Jy}$. (This refers to the usual {\it incoherent} self-absorbed
synchrotron flux; the black-body upper limits used for this estimate could
of course be violated by coherent emission mechanisms).

\noindent
a2) A slightly more involved impulsive model considers both the reverse and
the forward shock to be efficient radiators. An example of this occurs if for 
instance the ejecta has frozen-in magnetic fields, sufficiently strong to 
ensure radiation but not enough to dominate the dynamics (the FC model of 
RMP94, their Fig. 1a). At $t_\gamma\sim t_{dec}$ the reverse shock has achieved 
transrelativistic speed ($\bar{\Gamma_r} \sim 2$ respect to the ejecta, with 
$\kappa=10^3,~B'\sim 0.3 {\rm G}$ and $\Gamma\sim \eta\sim 10^3$ in this 
numerical example). The ejecta electrons have  $\gamma_{min}\sim 10^3$, giving 
an observed synchrotron peak at $\nu_m\sim 3\times 10^{14} \sim 10^{15}$ Hz 
with fluence $S_\nu \sim 10^{-8} K D_{28}^{-2}$ erg cm$^{-2} \propto \nu^0$ 
above $\nu_m$ (or flux $F_\nu \propto \nu^{-1}$), where $K=E_{51}
\theta_{-1}^{-2}$; 
then between $10^{18}$ and $3\times 10^{20}$ it has $S\nu\sim \nu^1$ (flux
$F_\nu \sim \nu^0$), and $S_\nu \sim 10^{-6} K D_{28}^{-2}$ erg cm$^{-2} 
\propto \nu^0$ ($F_\nu \propto \nu^{-1}$) above $3\times 10^{20}$ Hz, the 
latter two corresponding to IC-scattered synchrotron by reverse shock 
electrons.  We consider the evolution of the reverse shock (ejecta) region 
only, since it has stronger XR and O contributions than the forward blast wave.
(The forward blast wave remnant might be naively expected to behave as the a1 
case above, but being an IC component its contribution drops in time by two 
powers of $\gamma$ faster than the synchrotron component).  The shocked 
reverse gas is in pressure equilibrium with the gas forward of the CD,
so it also moves with $\Gamma\sim t^{-3/8}$ and $r\propto t^{1/4}$ as does 
the CD. The ejecta frozen-in field behaves as $B'\propto r^{-2}\propto 
t^{-1/2}$. The comoving pressure on either side of the side of the CD evolves 
$\propto r^{-3}$ so the reverse gas, even though traversed by reflected 
pressure waves, cools in its rest frame with $T'_p\propto r^{-6/5}$ and the 
comoving density in the ejecta evolves as $n'\propto r^{-6/5}\propto t^{-9/20}$,
the total number of particles in the ejecta remaining constant. The ejecta 
comoving radial width is $\Delta R' \propto r^{-1/5}$ and the ejecta column 
density is $n'.\Delta R' \propto r^{-2}\propto t^{-1/2}$.  As the ejecta cool,
any fresh acceleration of ejecta electrons can at most give $\gamma \propto 
\kappa T'_p \propto t^{-3/10}$, so the GRBR spectrum will be dominated by the
adiabatic cooling of the electron energy acquired in the initial GRB shock 
heating, $\gamma\propto r^{-1}\propto t^{-1/4}$. The reverse peak $\nu_m \sim 
10^{15} (t/t_\gamma)^{-11/8}$ Hz is produced by a comoving peak intensity
$I'_{\nu_m} \propto n'_e B' \Delta R' \sim 10^{-3} (t/t_\gamma)^{-1}$. 
Therefore $F_{\nu_m} \propto t^2 \Gamma^5 I'_{\nu_m} \sim 10^{-23} K 
D_{28}^{-2} (t/t_\gamma)^{-7/8}$ evolves proportional to $\nu_m^{7/11}$. The 
self-absorption frequency initially is $\nu_{ab}\sim 10^{13} 
(t/t_\gamma)^{-3/8}$ Hz, and the absorption and synchrotron peak frequencies 
become equal at $\nu_{m,ab}\sim 10^{11}$ Hz after a time $t_{m,ab}\sim 10^3 
t_\gamma$. Thereafter the peak of the self-absorbed flux occurs at a frequency 
$\nu_{ab} \sim 10^{11}(t/t_{m,ab})^{-57/56}$  where $F_{\nu_{ab}} \sim 3\times 
10^{-26} (\nu/\nu_{ab})^{69/57}$.  The optical flux near $10^{15}$ Hz is 
$F_{opt}\sim 10^{-23}(t/t_\gamma )^{-9/4}$ for $ t \geq t_\gamma$, both before 
and after $t_{m,ab}$.  (In principle one might have expected somewhat later 
a phase of $F_{opt} \sim$ constant due to the $\propto \nu^0$ IC component 
coming into the optical range, but as mentioned above, the IC energy losses 
drop off in time much faster than the synchrotron losses, so this should not 
be important). The visual magnitude of the GRBR from this reverse shock with 
frozen-in magnetic fields is therefore
\begin{equation}
m_v  \simeq  9 - 2.5 \log K + 5 \log D_{28} + (45/8) \log(t/t_\gamma) ~~~~ 
  \hbox{,~for} ~ t \simg t_\gamma ~.
\label{eq:a2mag}
\end{equation}
The incoherent flux in the 100 MHz radio band is maximum when the 
self-absorption peak passes through that frequency at the time $t_R \sim 
10^6 t_\gamma$, giving $F_{\nu_R ,max} \sim 1 K D_{28}^{-2} \mu {\rm Jy}$, 
which at 300 Mpc would be $\sim 0.1  K D_{27}^{-2}$ mJy, still very low.
Before the maximum of the radio flux (between $t_{m,ab}$ and $t_R$) the radio
flux grows as $F_{\nu_R} \propto t^{21/16}$, and after $t_R$ it decays as 
$F_{\nu_R} \sim t^{-9/4}$.

\noindent
a3) A somewhat similar impulsive model also considers both reverse and forward 
shock emission, but differs from (a2) in the origin of the magnetic field, 
which is here assumed on both sides of the contact discontinuity to be built 
up by turbulent motions to some fraction of the thermal proton energy density. 
The GRB spectrum corresponds to the TC model of RMP94 (their Fig. 1b). In this 
example the field energy is a fraction $\lambda \sim 10^{-6}$ of equipartition,
and all initial physical parameters (as well as the initial spectrum) are the 
same as in case a2) above; the optical peak is synchrotron from the
reverse shock. However, the magnetic field evolves in time differently,
${B'^2}/8\pi$ maintaining itself roughly at a constant fraction $\lambda$ of 
the thermal proton energy density $k n' T'_p \propto r^{-3}\propto t^{-3/4}$ , 
so $B'\propto t^{-3/8}$ in the reverse shocked ejecta. Retracing the steps in 
a2) we now have $I'_{\nu_m} \propto t^{-7/8}$, $\nu_m \propto t^{-5/4}$,
$F_{\nu_m} \propto t^{-3/4}$, or $F_{\nu_m} \propto \nu_m^{3/5}$ initially.
Also $\nu_{ab} \propto t^{-11/16}$, and $\nu_m$ becomes coincident with 
$\nu_{ab}$ at $\nu_{m,ab} \sim 10^{10.5}$ Hz at the time  $t_{m,ab} \sim 
3\times 10^3 t_\gamma$, where the peak value is $F_{\nu_{ab}} \sim 
3\times 10^{-26}$. After that time the peak of the self-absorbed spectrum 
moves down as $F_{\nu_{ab}} \propto \nu_{ab}^{59/53}$, with $\nu_{ab} 
\propto t^{-53/56}$. The optical flux around $10^{15}$ Hz is
$F_{opt} \sim 10^{-23} K D_{28}^{-2} (t/t_\gamma)^{-2} {\rm erg~cm^{-2}~s^{-1}~
Hz^{-1}}$ both before and after $t_{m,ab}$, so the visual magnitude of the
GRBR is
\begin{equation}
m_v  \simeq  9 - 2.5 \log K + 5 \log D_{28} + 5 \log(t/t_\gamma) ~~~~ 
  \hbox{,~for} ~  t \simg t_\gamma ~.
\label{eq:a3mag}
\end{equation}
The self-absorption peak comes into the 100 MHz radio range at $t_R \sim 
3\times 10^2 t_{m,ab} \sim 10^6 t_\gamma$ (a few weeks
after the GRB, for fiducial parameters with $t_\gamma\sim 1 {\rm s}$), and the 
peak radio flux is $F_{\nu_R , max} \sim 10^{-2} K D_{28}^{-2}$ mJy, or 
more interestingly, $F_{\nu_R ,max} \sim 1 K D_{27}^{-2}$ mJy for a GRBR at 
300 Mpc.  Between $t_{m,ab}$ and $t_R$ the radio flux grows as $F_{\nu_R} 
\propto t^{21/16}$, and decays as $F_{\nu_R} \sim t^{-2}$ after $t_R$.

\section{GRBR Spectra from Wind Fireball Models}

For wind models, GRB spectra have been discussed, e.g. by ~\cite{mr94} and ~
\cite{thom94,thom96}. The observed fluxes from the resulting GRBR will be 
given as before by equations (1) and (2), but there are significant 
differences from the impulsive case due to the smaller radii at which shocks 
occur, the different evolution dynamics (\S 2, b) and the different post-shock 
behavior of the physical quantities. Below we consider two particular examples 
of wind internal shock models.

\noindent
b1) The simplest wind GRB spectrum would be one where the break around 
$10^{20} {\rm Hz} \sim 0.5 {\rm MeV}$ is due to the synchrotron peak, e.g., a 
case resembling the (a1) impulsive spectra. We take as a specific example 
(\cite{pap96}), a case with $\eta=10^2,~t_w=10^2 {\rm s},~t_{var}\sim
2\times 10^{-2}{\rm s}$, and at $r_{dis}\sim 6\times 10^{12} t_{var,-2}\eta_2^2
{\rm cm}$ the turbulently generated field is $B'\sim 10^6$ G and
$\kappa\sim 10^3$ (internal shocks have $\Gamma_{rel}\sim 1$, so $\gamma_{min}
\sim 10^3$ and $\nu_m\sim \eta \nu'_m \sim 10^{20}$ Hz. The total energy 
into $\theta\sim 10^{-1}\theta_{-1}$ is again $E=10^{51}E_{51}$ erg spread over
$t_w=10^2 t_{w,2}{\rm s}$.  The GRB spectrum has the same shape and fluence
as the impulsive case (a1), but the flux level is $10^{-2}$ because the 
duration is $10^2$ longer in this example. We consider the GRBR regime for
$r \simg r_{imp}\sim 3\times 10^{16} t_{w2}\eta_2^2 {\rm cm}$. However, in this 
type of model the comoving $t'_{sync}\sim 5\times 10^{-7} B_6^{-2} 
\gamma_{3,min}^{-1} \ll t'_{ex}$, and the remnant undergoes essentially 
instantaneous cooling as soon as shocks stop heating the gas. This is 
characteristic of a wind model with a GRB break at 0.5 MeV ascribed to a 
synchrotron peak, since this requires high $B'$ and $\gamma$. So in these cases 
we expect no significant emission from the GRBR after the gamma-ray flash, 
although there will be (as in a1) some contemporaneous optical and X-ray 
emission during the GRB itself, for $t < t_w$.

\noindent
b2) Consider now GRB from wind internal shocks with a synchrotron peak at
optical frequencies and an IC peak around 0.5 MeV. As a specific example 
(\cite{pap96}) we take a model with $t_w\sim 10^2{\rm s},~t_{var}\sim 
2\times 10^{-2}{\rm s},~ \kappa=10^3,~\eta=10^2$, and at $r_{dis}\sim 6\times 
10^{12} t_{var,-2}\eta_2^2 {\rm cm}$ we assume that magnetic fields build up 
by turbulent motions to a fraction $\lambda\sim 10^{-6.5}$ of the proton 
thermal energy, or $B'\sim 3$ G,  which leads then to a GRB synchrotron optical 
peak at $\eta \nu'_m \sim 10^{15}$ Hz, and an IC peak near $10^{20}{\rm Hz}\sim 
0.5$ MeV. The GRB fluence spectrum $\nu F_\nu t_w$ is the same as for the
impulsive models b2and b3, and the flux has the same shape but is a factor 
$10^{-2}$ lower since here $t_w \sim 10^2$ s. In this case, the comoving 
synchrotron timescale $t'_{sy} \sim 5\times 10^3 \gg t'_{ex}$, so the adiabatic 
approximation will be valid in the remnant, which retains a significant 
thermal energy content that is radiated away during the later expansion. 
During the GRB the initial optical peak at $\nu_m\sim 10^{15} {\rm Hz}$ has 
flux $F_{\nu_m}\sim 10^{-25} t_{w2}^{-1} K D_{28}^{-2}$ erg cm$^{_2}$ s$^{-1}$ 
Hz$^{-1}$. The comoving intensity is, from equation (1), $I'_{\nu_m}= F_{\nu_m} 
D^2/ (c^2 t^2 \Gamma^5 )\sim 10^{-4} K D_{28}^2 t_{w2}^{-1}\eta_2^{-5}$ and the 
self-absorption frequency is $\Gamma (I_{\nu'} c^2/ [2 \gamma m_e c^2] )^{1/2} 
\sim 10^{12}{\rm Hz}$. For the time scaling we use the properties of the wind 
in the impulsive $\Gamma\simeq$ constant phase (\ref{eq:windgamma}), so the 
comoving synchrotron intensity scales as $I'_{\nu_m} \sim n'_e \Delta R' B' 
\propto t^{-4}$, the synchrotron peak frequency as $\nu_m \sim \Gamma B' 
\gamma^2 \propto t^{-4}$, and the observed flux at the synchrotron peak as 
$F_{\nu_m} \sim c^2 t^2 D^{_2} \Gamma^5 I'_{\nu_m} \propto t^{-2}$, so that 
the synchrotron peak moves down with peak frequency as $F_{\nu_m} \propto
\nu_m^{1/2}$. The self-absorption frequency is $\nu_{ab}\propto \Gamma (I'_\nu
/\gamma)^{1/2} \propto t^{-3/2}$, and the synchrotron maximum and 
the self-absorption frequencies coincide at $t_{m,ab}\sim 10^{6/5}t_w$ 
at $\nu_{m,ab}\sim 2\times 10^{10}$ Hz, where the flux is
$F_{\nu_{m,ab}}\sim 10^{-27}$. After that the self-absorbed peak of
the spectrum moves down in frequency as $F_{nu_{ab}} \propto \nu_{ab}^{4/3}$.
The optical flux as a function of time is given by $F_{opt}\sim
F_{\nu_m}(\nu_m/10^{15}) \sim 10^{-25} K D_{28}^{-2} (t/t_w)^{-6}$. 
The corresponding GRBR visual magnitude is, as a function of time,
\begin{equation}
m_v \simeq 14 -2.5\log K +5\log D_{28} +15\log(t/t_w) ~~~~~\hbox{,~for} t> t_w~,
\end{equation}
where $K=E_{51}\theta_{-1}^{-2} t_{w2}^{-1}$. The initial magnitude (14,
in this example) is constant for $t \siml t_w=10^2$ s during the GRB, and 
it drops extremely fast in the GRBR phase afterwards. AT 300 Mpc, the initial 
magnitude is $m_v \sim 9$, dropping to 24 after $t\sim 10 t_w \sim 10^3$ s
for these parameters.  The radio flux in this model is very low, $F_R \sim 
0.3 K D_{28}^{-2} \mu {\rm Jy}$, or $0.03 K D_{28}^{-2}$ mJy at 300 Mpc,
growing as $t^3$ before $t_R \sim 3 \times 10^2 t_w \sim 3\times 10^4 {\rm s}$ 
in this example, and dropping as $t^{-6}$ afterwards.

\section{Discussion and Observational Prospects}

Gamma-ray burst remnants, or GRBR, should according to our calculations 
leave behind an afterglow at wavelengths longer than $\gamma$-rays, in 
particular at X-ray, optical, and in some cases also in the radio bands.
The calculations discussed above apply both to spherically expanding 
configurations and to jet configurations, as long as the jet opening
angle $\theta \simg \Gamma^{-1}$. The adiabatic approximation made
for the evolution of the remnant should be generally valid for times 
significantly longer than the gamma-ray burst duration. A wide variety of 
models, involving several unknown parameters, are compatible with the gamma-ray
data. Observations in other bands offer the chance to narrow the range of
options and refine existig models.  The numerical examples discussed are 
illustrative of the possibilities, and should give a reasonable idea of the 
range of values that might be expected. There is greater reason for confidence 
in the scaling laws discussed, and in particular in the time dependence laws 
of the fluxes in different regimes for the various models, as these are based 
on simple physical arguments. 

Omnidirectional, or at any rate large field-of-view X-ray detectors
in space (such as HETE, ~\cite{ri92}) may be able to detect the simultaneous
X-ray emission predicted for GRB (e.g. MRP94,\cite{pap96}) as well as,
for the brighter bursts, possibly the X-ray afterglow implied by our
models discussed here, although surface areas higher than HETE's may
be required to follow the remnant evolution. 
%From the point of view
%of availability of significant collecting areas and fast response time,
%optical observations may stand a much better chance of detecting GRBR.

The synchrotron radio fluxes from GRBR are expected to be very small, reaching 
their maximum value on timescales of weeks after the GRB outburst, which in the
optimal cases (e.g. \S 2, b3) may be at most in the mJy range. The radio flux 
is expected to change before and after the maximum as power laws of the time 
which are characteristic of the models. Nonetheless, radio searches, even 
for very short timescale flashes, are worthwhile because we cannot rule out
coherent emission behind relativistic shocks, which would of course permit
brightness temperatures far higher than the usual self-absorption limit.
(indeed, the intraday radio variations in AGN [\cite{wa96}] may exemplify
coherent emission from shocks whose properties resemble those expected in
GRBRs).

The optical detection of the GRB event itself is within the range of
capabilities of modest size telescopes, but the problem is one of field
of view, as typical BACODINE error boxes supplied within a minute of the GRB
event are at least several degrees, and sometimes as much as ten degrees 
wide, and any significant improvement in location takes at least days. Thus
detection of optical emission from the burst itself (rarely longer than minutes)
is difficult. However the GRBR optical emission decays on longer timescales,
typically hours, and for the initial magnitude levels implied by several of 
the models even at Gpc distances (e.g. $m_v \sim 9-14$, see \S3 a2, a3, 
\S 4 b2), meter class telescopes equipped with CCD detectors may be able
to cover a several square degree wide field by doing rasters of ten minute 
observations. The advantage of such observations, if they can be repeated
more than once over the same field, is that they would allow a determination 
of the optical flux time-decay exponent, which can help discriminate
among models.

\acknowledgements
We are grateful to NASA NAG5-2857 and the Royal Society for support.

\end{document}